\documentclass[aps,prl,twocolumn,superscriptaddress,showpacs]{revtex4}
\usepackage{graphicx}
\begin{document}
%\baselineskip 24pt
%\draft
\title{Dependence of the $^{12}$C($\vec{\gamma}$,pd) reaction on photon linear polarisation
\\}
                                                                                
%\author{D. P. Watts$^1$ etc..}
%\address{Dept. of Physics and Astronomy}

\author{D. P. Watts}
\email{daniel.watts@ed.ac.uk}
\altaffiliation[Present address]{ Edinburgh University, UK.}
\author{J.R.M. Annand}\affiliation{Department of Physics and Astronomy, University of Glasgow, Glasgow, G12 8QQ, UK.}
\author{R. Beck}
\affiliation{Institut f\"{u}r Kernphysik, Universit\"{a}t Mainz, D-55099 Mainz,
Germany.}
\author{D. Branford}
\affiliation{School of Physics, Edinburgh University, EH9 3JZ, UK.}
\author{D. Glazier}
\author{P. Grabmayr}
\affiliation{Physikalisches Institut, Universit\"{a}t T\"{u}bingen, D-72076 T\"{u}bingen,Germany.}
\author{K. Livingston}
\author{I. J. D. MacGregor}
\author{J.C. McGeorge}
\author{R.O. Owens}
\affiliation{Department of Physics and Astronomy, University of Glasgow, Glasgow, G12 8QQ, UK.}
%D. Branford$^2$,  D. Glazier$^{1}$, P. Grabmayr$^3$, \\K. Livingston$^1$, 
% I. J. D. MacGregor$^1$, J.C. McGeorge$^1$, R.O. Owens$^1$}
%\address{$^1$Department of Physics and Astronomy, University of Glasgow, Glasgow G12 8QQ, UK\\
%$^2$School of Physics, University of Edinburgh, Edinburgh EH9 3JZ, UK\\
%$^3$Physikalisches Institut, Universit\"{a}t T\"{u}bingen, D-72076 T\"{u}bingen,Germany\\
%$^4$Institut f\"{u}r Kernphysik, Universit\"{a}t Mainz, D-55099 Mainz,
%Germany\\}

\date{\today}

\begin{abstract}
The sensitivity of the $^{12}$C$(\vec{\gamma},pd)$ reaction to photon linear polarisation has been determined at MAMI, giving the first measurement of the reaction for a nucleus heavier than $^{3}$He. Photon asymmetries and cross sections were measured for $E_{\gamma}$=170 to 350 MeV. For $E_{\gamma}$ below the $\Delta$ resonance, reactions leaving the residual $^{9}$Be near its ground state show a positive asymmetry of up to 0.3,  similar to that observed for $^{3}$He suggesting a similar reaction mechanism for the two nuclei.

\end{abstract}

\maketitle

%\narrowtext

%\section{Introduction}

Three-body forces have consequences in many fields of physics. The study of photon induced proton-deuteron knockout from nuclei may give valuable information on the three-body interaction in the nucleus, since the direct mechanisms which contribute may be related to those thought to be involved in the three-nucleon force\cite{Skibinski:2002ak,Skibinski:2003de,Deltuva:2003xe,Laget:1988hv}. However, as well as the direct 3-nucleon process ($3N$) there will be contributions from initial photon absorption by a single nucleon ($1N$), two-nucleons ($2N$) and two-step $3N$ processes such as initial real pion production on one nucleon followed by reabsorption by a nucleon pair. Clearly, to extract reliable information from  $(\gamma,pd)$ measurements, the relative contributions from each of these mechanisms should be well understood.

The ($\gamma$,pd) reaction has received significant theoretical interest in recent years, mainly motivated by the possibility of obtaining information on the nature of the 3-nucleon force (3NF). Detailed $^{3}$He calculations based on exact solutions of the three-particle scattering equations in the initial and final states have been carried out for photon energies up to 140 MeV\cite{Skibinski:2002ak,Deltuva:2003xe}. These show that the inclusion of a 3NF has a large effect on the magnitude of the cross section, increasing the predictions by up to a factor of two at the top end of this $E_{\gamma}$ range. A microscopic theoretical treatment of the $^{3}$He($\gamma$,pd) reaction, which includes contributions from $1N$, $2N$ and $3N$  mechanisms, has been developed by Laget\cite{Laget:1988hv}. The $3N$ mechanisms include contributions from virtual and real pion exchange. Laget's treatment relies on a factorisation approximation to simplify the computation but is applicable up to higher photon energies.

On the experimental side, most measurements of the ($\gamma$,pd) reaction have been made using $^{3}$He targets\cite{Isbert:1993xf,gpd2,argan,Gassen:1981ek,gpd5,kolb}. An important feature of the measured excitation functions is that they show no evidence of structure for photon energies in the $\Delta$(1232) resonance region. Also, the centre-of-momentum (CM) proton angle distributions are forward peaked and fall off rapidly with increasing angle up to $\sim$70$^{\circ}$ with a flatter distribution at more backward angles. The features are moderately well described by the Laget model when $1N$, $2N$ and two-step $3N$ (including only real $\pi$ exchange) mechanisms are included\cite{Laget:1988hv}. Laget notes that his model accounts less well for the $^{3}$He$(\gamma,pd)$ data than it does for $\pi$ induced processes involving the A=3 nuclei and suggests two additional photon couplings, both involving two highly virtual mesons, which could be responsible. Above 100 MeV the $2N$ and two-step $3N$ mechanisms are predicted to dominate with the $2N$ mechanisms only giving large contributions to the cross section at forward CM proton angles\cite{Laget:1988hv,Isbert:1993xf}. Above $\sim$150~MeV the two-step 3N mechanism is predicted to provide most of the cross section for CM proton angles backwards of $\sim$70$^{\circ}$.

Studies of the $(\gamma,pd)$ reaction for A$>$3 targets have been carried out only on $^{16}$O\cite{Hartmann:1973} and $^{12}$C\cite{McAllister:1999mi}. Both measurements show a photon energy dependence similar to that observed in $^{3}$He with no prominent enhancement for $E_{\gamma}$ around the $\Delta$ resonance. This behaviour is in contrast to photon induced $pp$, $pn$, $p\pi$ and $ppn$\cite{MacGregor:1998,Branford:1999cp,Audit:1996tq,Watts:2002jt} knockout reactions where the $\Delta$ plays a prominent role. The  $^{12}$C($\gamma,pd)$ missing energy spectra obtained in Ref. \cite{McAllister:1999mi} exhibit significant strength close to the reaction threshold, and the recoil momentum spectra of the (A--3) nucleus at low missing energies are consistent with those predicted if it were a spectator to the knockout of three 1p-shell nucleons.

The photon asymmetry for the $(\vec{\gamma},pd)$ reaction has only been measured previously for $^{3}$He\cite{Belyaev:1985kg,fabbri:1972}. These measurements showed a positive asymmetry which ranged from around 0.2 to 0.5 over the sampled photon energy region of 90-350~MeV and CM proton angle range of 60-135$^{\circ}$. Comparison with a simple Faddeev calculation\cite{Belyaev:1985kg} which neglects meson exchange currents, $\Delta$ contributions and final state interactions gave limited agreement with the experimental data.

The three nucleon photoabsorption mechanisms which operate in the  $(\gamma,pd)$ reaction also contribute in the $(\gamma,ppn)$ reaction but with less restrictive spin and isospin conditions for the final state particles\cite{Carrasco:1992mg,Audit:1996tq}. Recent Faddeev theoretical calculations for $^{3}$He\cite{Skibinski:2002ak} indicate that the $(\gamma,ppn)$ reaction is also sensitive to the nature of the 3NF. Several measurements of this reaction have been made in the last decade\cite{ruth:1994,Audit:1996tq,Watts:2002jt,Sarty:1993}. Polarised photon measurements of the $^{3}$He$(\vec{\gamma},p)X$ reaction\cite{ruth:1994} for regions of nucleon momenta where the two-step 3N processes were expected to dominate showed a negative asymmetry of up to 0.2 which is not well described by the Laget model. A recent comparison of the $^{12}$C$(\gamma,ppn)$ reaction\cite{Watts:2002jt}, with model calculations\cite{Carrasco:1992mg} in restricted kinematics for the detected nucleons gave clear evidence for the existence of a direct $3N$ mode.  

The present work is the first measurement of the $^{12}$C($\vec{\gamma}$,pd) reaction. The sensitivity of the cross section to photon linear polarisation is expected to give valuable constraints on the reaction mechanisms for different photon energies and in different excitation energy regions of the (A--3) nucleus. 

%\section{EXPERIMENT.}
The experiment was carried out at the 855~MeV Mainz microtron (MAMI-B)
\cite{mamib,mamib2} using the Glasgow 
tagged-photon spectrometer\cite{Anthony:1986ge,Hall:1996gh} in conjunction with two plastic scintillator
arrays, PiP and TOF\cite{MacGregor:1996tb,Grabmayr:1998ez}, set up as described in Ref. \cite{Powrie:2001af,Franczuk:1999}. Polarised photons were produced by coherent bremsstrahlung in a thin diamond radiator\cite{Lohmann:1994vz,Kraus:1997pp,Rambo:1998iu,Natter:2002}. The polarisation orientation of the photons was flipped between horizontal and vertical every few minutes. Three angular settings of the diamond were used for which the main coherent peak covered the photon energy ranges 170--220, 220--280 and 300--350 MeV with corresponding average linear polarisation ($P$) of 59.5\%, 49.5\% and 42.5\% respectively. 

Protons with kinetic energies 31-270 MeV were detected in the charged particle hodoscope PiP covering the polar angular 
 range $\theta$=51$^{\circ}$-129$^{\circ}$ and azimuthal angular range of $\phi$=$\pm$23$^{\circ}$. 
%The lower threshold was
%determined by the proton energy losses before entering PiP and its 8~MeV pulse height threshold. High energy
%protons which entered the back layer of PiP were vetoed and a software cut was
%applied to the remaining proton events to give a well determined maximum 
%energy.
%The reaction timing was obtained from a 
%segmented half ring of 1~mm thick scintillators ($\Delta$E$_{\text{PiP}}$) centred on the target
%and positioned on the PiP side of the photon beam at a radius of $\sim$11~cm. 
%The first level trigger required coincident hits in a PiP element and a 
%$\Delta$E$_{\text{PiP}}$ element between PiP and the target.
Coincident deuterons leaving the target with energies above  $\sim$45~MeV were detected in TOF which determined particle energies by time-of-flight. The TOF detectors covered $\theta$=10.0$^{\circ}$-175.0$^{\circ}$. 
%Charged particles in TOF were selected using information from two segmented half rings of scintillator ($\Delta$E1, $\Delta$E2), each 2~mm thick, centred on the target at radii of $\sim$11~cm and $\sim$30~cm. 
The deuterons were separated from other charged hadrons in TOF by selecting events from a 2-D plot of inverse speed versus pulse height, as described in Ref. \cite{McAllister:1999mi}. The number of random deuteron coincidences in TOF was found to be negligible. The average measured 
missing energy resolution extracted from  D$(\gamma,pn)$ was found to be $\sim$7~MeV. For $(\gamma,pd)$ the average resolution is better due to the slower flight times for the heavier particles and is estimated to be $\sim$5 MeV. The $\sim$3\% background of events not originating from reactions in the target was measured in runs with the target removed. The total systematic error in the asymmetry is estimated to be $\Delta \Sigma$ = $\pm$0.05$\Sigma$\cite{Powrie:2001af}. The systematic uncertainties in the measured cross sections are estimated to be up to $\pm$8\%\cite{McAllister:1999mi}.
 
For each goniometer setting all events due to photons in the coherent peak region were used to produce average $(\vec{\gamma},pd)$ cross sections ($\sigma$) and asymmetries ($\Sigma$) for the events which were within the geometrical and energy acceptances of the PiP-TOF detector systems. The asymmetry is defined as $\Sigma = \frac{1}{P}\frac{\sigma^{\parallel}-\sigma^{\perp}}{\sigma^{\parallel}+\sigma{\perp}}$  where $\sigma^{\parallel}$ ($\sigma^{\perp}$) is the measured cross section for reactions in which the horizontal detector plane is parallel(perpendicular) to the electric vector of the polarised photons. The reduction of the asymmetry arising from the particle detectors having a finite $\phi$ acceptance around the horizontal plane is estimated to be $\sim$10\%. A further reduction which arises from the smearing of the photon polarisation direction in the (3N+$\gamma$) CM frame due to the initial 3N Fermi momentum is estimated to be $\sim$5\%. The magnitude of the presented asymmetries have been increased by 15\% to account for these effects.
%\section{RESULTS}

The cross section as a function of $E_{m}$ is shown in Fig.~1 for three different $E_{\gamma}$ bins.  Missing energy is defined as $E_{m}~=~E_{\gamma}~-~T_{1}~-~T_{2}~-~T_{r}$ where $E_{\gamma}$ is the incident photon energy, 
$T_{1}$ and $T_{2}$ are the energies of the two detected particles and $T_{r}$ is the (typically small) energy of the recoiling system which is calculated
from its momentum ${\bf P}_{r}={\bf P}_{\gamma}-{\bf P}_{p}-{\bf P}_{d}$.  The Q-value for the $^{12}$C$(\vec{\gamma},pd)$ reaction is --31.7 MeV. The missing energy spectra show strength near to the Q-value and also a peak at $\sim$45 MeV. Above this peak, the $(\gamma,pd)$ missing energy distribution rises to a second maximum, which becomes higher and wider as $E_{\gamma}$ increases. Similar general features in the missing energy are seen in the earlier work of Ref. \cite{McAllister:1999mi}, but with weaker indications of the peak at $\sim$45 MeV, probably due to its poorer statistical accuracy and inferior resolution. A $(\gamma,pd)$ reaction which involves the knockout of three (1p) shell nucleons leaving a residual (A--3) spectator nucleus would populate missing energies up to $\sim$50 MeV. The missing energies expected if the residual $^{9}$Be is left in one of its four low lying and relatively long lived states are also indicated in Fig.~1. The peak at $\sim$45 MeV is seen to occur at energies close to the lowest lying T=$\frac{3}{2}$ states at 14.4 and 17.0 MeV.  Also shown in Fig.~1 are the scaled $^{12}$C$(\gamma,pp)$ and $^{12}$C$(\gamma,pn)$ cross sections obtained with the same detector setup as the present measurement (this data was already analysed in Ref. \cite{Powrie:2001af}).  Above $E_{m}\sim$60 MeV the $(\gamma,NN)$ reactions scale with $E_{m}$ and $E_{\gamma}$ in a similar way to the present $^{12}$C$(\gamma,pd)$ data. At low $E_{m}$ the channels show different behaviour. The $(\gamma,pp)$ data falls off more rapidly than $(\gamma,pd)$ and the $(\gamma,pn)$ cross section has a large peak  at $E_{m}\sim30$MeV (off scale in Fig.~1) due to a large direct two-nucleon knockout contribution.
% The 0.0 and 2.4 MeV states both have  T=$\frac{1}{2}$, and those at 14.4 and 17.0 MeV are the two lowest lying T=$\frac{3}{2}$ states.

\begin{figure}
\includegraphics[width=80mm]{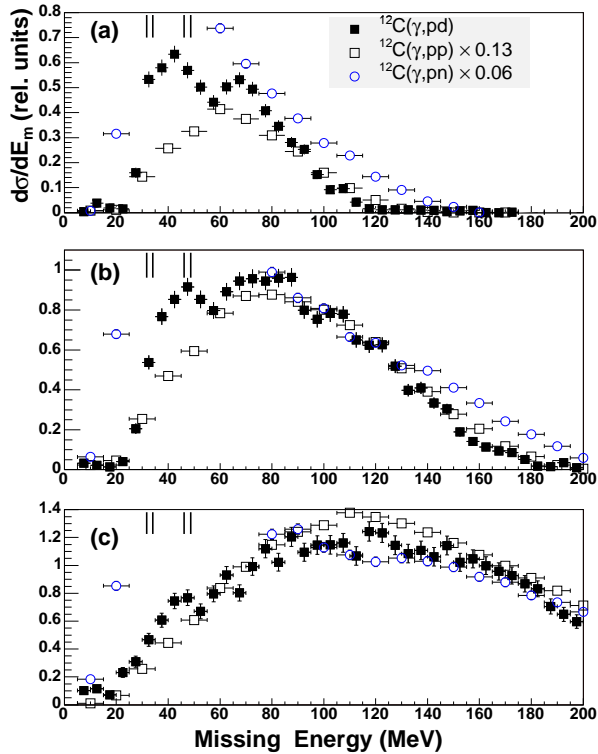}
\caption{Cross sections for the ($\gamma,pd$), ($\gamma,pp$) and ($\gamma,pn$) reactions versus missing energy are shown as the filled squares, open squares and open circles, respectively. The data are presented for $E_{\gamma}$ bins of (a) 170-220~MeV, (b) 220-280~MeV and (c) 280-350~MeV. The  ($\gamma,pp$) and ($\gamma,pn$) cross sections have been normalised by the factors indicated on the figure. The ground state and the three narrow excited states in $^{9}$Be at 2.4, 14.4 and 17.0 MeV are indicated by the marks on  the top of each panel.
\label{fig1}}
\end{figure}

%\begin{figure}
%\includegraphics*[0mm,0mm][220mm,255mm]{gpd_fig1_10.eps}
%\caption{Cross sections for the ($\gamma,pd$), ($\gamma,pp$) and ($\gamma,pn$) reactions versus missing energy are shown as the filled squares, open squares and open circles, respectively. The data are presented for $E_{\gamma}$ bins of (a) 170-220~MeV, (b) 220-280~MeV and (c) 280-350~MeV. The  ($\gamma,pp$) and ($\gamma,pn$) cross sections have been normalised by the factors indicated on the figure. The ground state and the three narrow excited states in $^{9}$Be at 2.4, 14.4 and 17.0 MeV are indicated by the marks on  the top of each panel.
%\label{fig1}}
%\end{figure}

The photon asymmetry for the $^{12}$C$(\gamma,pd)$ reaction is presented in Fig.~2. The $E_{m}<$40 MeV region emphasizes $(\gamma,pd)$ reactions leading to the ground state and low lying excited states  of $^9$Be, which are  all T=$\frac{1}{2}$. A positive asymmetry is observed for photon energies up to $E_{\gamma}\sim$280 MeV, while at higher photon energies in the  $\Delta$ resonance region the asymmetry becomes negative. The asymmetry for $^{3}$He($\vec{\gamma},pd$) for $\theta_{p}$=90 and 110$^{\circ}$ in the CM frame\cite{Belyaev:1985kg} (both corresponding to lab proton angles covered by the PiP  detector in the present measurement) are also shown in Fig.~2(a). The magnitude and sign of the $^{3}$He asymmetry is similar to the $E_{m}<$40 MeV $^{12}$C data for photon energies up to $\sim$270 MeV. At higher $E_{\gamma}$ the $^{3}$He data do not show the negative or small asymmetries indicated in the $^{12}$C($\gamma,pd$) data.

The $E_{m}$=40-50 MeV cut (Fig.~2(b)) emphasizes $(1p)^{3}$ knockout events leading to higher excited states and includes the peak region visible in the cross section as a function of missing energy (Fig.~1).  The asymmetry in this region is generally negative or small below $\sim$300 MeV in contrast to the positive asymmetry observed at lower missing energies for both $^{12}$C and $^{3}$He($\vec{\gamma},pd)$. The asymmetry for the $E_{m}$=50-100 MeV region shows the same general trends as the $E_{m}$=40-50 MeV data, albeit smaller in magnitude. The asymmetries for both these higher $E_{m}$ regions can be seen to show features which are very similar to those observed for the $^{12}$C($\vec{\gamma},NN$) reactions at high missing energy\cite{Powrie:2001af}, suggestive of similar underlying reaction mechanisms. The model of Ref. \cite{Carrasco:1992mg} explains the $(\gamma,NN)$ cross section in this missing energy region largely as the result of detecting two of the three (or more) nucleons produced by a two-step 3N process or by initial photon absorption on a two-nucleon pair followed by final state interactions\cite{Lamparter:1996va,Watts:2000dk}.  The same processes, where one of the outgoing nucleons picks up an additional nucleon from the residual nucleus, can probably explain the similar $E_{m}$ distribution and asymmetry of the $^{12}$C$(\gamma,pd)$ reaction. As these mechanisms involve more than three nucleons they have no  analogue in the reaction on $^{3}$He.

A comparison of the $(\gamma,pp)$ and $(\gamma,pd)$ $E_{m}$ distributions below 60 MeV may suggest the pickup process discussed in  the previous paragraph still provides a  significant background contribution in this region. However, this extrapolation should be considered  an upper limit as the dominant mechanism of the $(\gamma,pp)$ reaction changes to direct  two-proton emission following photon absorption on two-nucleons\cite{Lamparter:1996va,Watts:2000dk} at low $E_{m}$, which only has significant subsequent pickup probability at more forward proton angles than sampled here\cite{Laget:1988hv}. This is supported by the comparison in Fig.~2 of the $^{12}$C$(\vec{\gamma},pd)$ asymmetry with the asymmetry of the $^{12}$C($\vec{\gamma},pp$), and $^{12}$C($\vec{\gamma},pn$) reactions for $E_{m}\leq$40 MeV\cite{Powrie:2001af}. Both the reactions show a negative asymmetry.  The positive asymmetries observed in  $^{12}$C($\vec{\gamma},pd$) therefore argue against a large direct feeding of strength from ($\gamma,pp$) and ($\gamma,pn$) reactions through subsequent pickup reactions at the lowest missing energies.
 
%Figure 1 shows the separate $\sigma^{\parallel}$ and $\sigma^{\perp}$ components as  a function of missing energy and  Fig. 2 gives the photon energy dependence of the calculated asymmetry. The data in Fig. 1 indicate that for lower $E_{\gamma}$ the two photon polarisations tend to produce different missing energy distributions, with $\sigma^{\parallel}$ more preferentially populating the lower missing energies and $\sigma^{\perp}$ showing a stronger peak at $\sim$45 MeV. The $E_{\gamma}$ dependence of the $^{12}$C($\vec{\gamma},pd$) asymmetry, calculated from $\sigma^{\parallel}$ and $\sigma^{\perp}$, is presented in Fig. 2 
The selection of $^{12}$C$(\vec{\gamma},pd)$ events with low missing energy should enhance the contribution of processes which involve only the three detected nucleons while the (A--3) nucleus spectates. The similar asymmetry observed for $(\gamma,pd)$ reaction in $^{12}$C  in this $E_{m}$ region and $^{3}$He (Fig.~2(a)) suggests similar reaction mechanisms in both nuclei. The Laget model predicts that the two-step 3N mechanism involving the initial production of an on-shell pion is the dominant mechanism for the $E_{\gamma}$ and $\theta_{p}$ sampled in the present experiment. The published Laget calculations for $^{3}$He\cite{Laget:1988hv} do not present results for the ($\gamma,pd$) asymmetry, but an indication of its behaviour is sought here by examining the asymmetry in the initial p($\gamma,N$)$\pi$ stage of the dominant 3N two-step mechanism. Since the $(\gamma,pd)$ process involves deuteron formation from the recoiling nucleon in this process, then the underlying ($\gamma,N$)$\pi$ asymmetry should be reflected in the $(\gamma,pd)$ data. This feeding of the underlying ($\gamma,N$)$\pi$ asymmetry to the final state particles in the two-step 3N mechanism was already indicated in the $(\gamma,NN)$ measurements\cite{Powrie:2001af} at high missing energy. 

To determine whether these considerations apply to our data we used the predictions for the $\vec{\gamma} N \rightarrow N\pi$ asymmetry obtained using the MAID code\cite{maid}, which is based on a unitary isobar parameterisation of experimental data and accounts for the non-resonant and resonant parts of the pion photoproduction amplitude. Both  p($\vec{\gamma},\pi^{+})$n and  p($\vec{\gamma},\pi^{0})$p were calculated at a pion CM breakup angle of 55$^{\circ}$, which for photon energies above 200 MeV results in recoiling nucleon angles in regions where the deuteron yield in the present $^{12}$C$(\vec{\gamma},pd)$ data is largest ($\sim$40-60$^{\circ}$). The predictions are presented in Fig.~2(a).
\begin{figure}
%\rotatebox{0}{
%\resizebox{0.55\textwidth}{!}
\includegraphics[width=90mm]{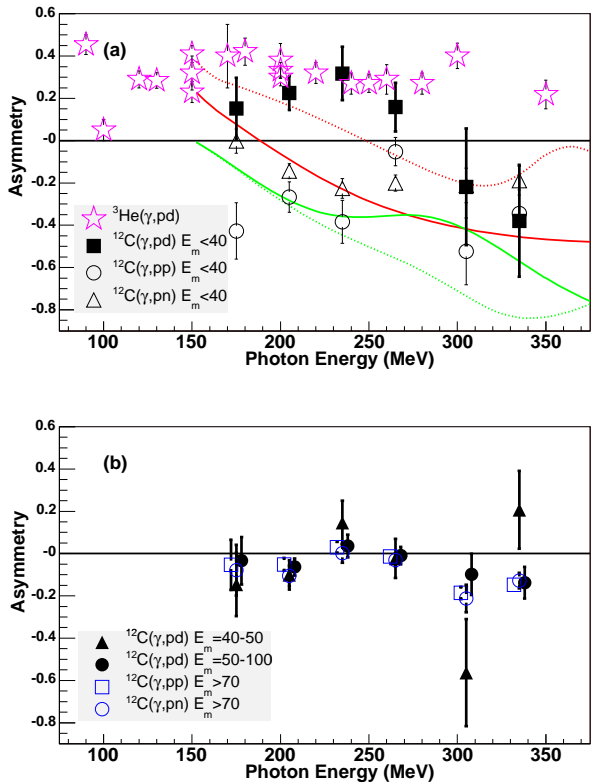}
\caption{(a) The $^{12}$C($\vec{\gamma},pd$), $^{12}$C($\vec{\gamma},pn$) and $^{12}$C($\vec{\gamma},pp$) photon asymmetry are presented for the missing energy cuts indicated in the figure. The results from $^{3}$He($\vec{\gamma},pd$) for $\theta_{p}$=90$^{\circ}$ and 110$^{\circ}$ are also shown. The  solid red and green lines give the asymmetry at $\theta_{\pi}$=55$^{\circ}$ from the MAID calculations for p($\vec{\gamma},\pi^{0}$)p and  p($\vec{\gamma},\pi^{+}$)n respectively. The corresponding predictions without including the $\Delta(1232)$ are shown by the dotted lines. (b) Comparison of the $^{12}$C($\vec{\gamma},pd$), $^{12}$C($\vec{\gamma},pn$) and $^{12}$C($\vec{\gamma},pp$) asymmetry at higher missing energy. \label{fig2}}
\end{figure}
Neither of the full p($\gamma,\pi$)$N$ calculations can give a simple explanation of the low missing energy $^{12}$C$(\vec{\gamma},pd)$ asymmetry in terms of the two-step 3N process. As  the $\Delta$ contribution to the initial pion production  vertex for the two-step 3N process is suppressed due to an isospin restriction\cite{Laget:1988hv}, MAID calculations with the $\Delta$ contribution removed are also shown in Fig.~2. While the  p($\gamma,\pi^{0}$)p MAID prediction with no $\Delta$ contribution comes closer to the $(\vec{\gamma},pd)$ asymmetry, the MAID cross sections suggest a dominance for the p($\gamma,\pi^{+}$)n process for which the asymmetry is negative. It therefore seems unlikely that there is  a simple explanation  of the  $^{12}$C$(\vec{\gamma},pd)$ or  $^{3}$He$(\vec{\gamma},pd)$ asymmetries in terms of the two-step 3N mechanism. 

It is clear that theoretical input is required to fully account for the various mechanisms and their possible interference. Of particular interest would be estimates which include 3N mechanisms involving heavier mesons and also the two-meson couplings suggested by Laget. Such processes produce a shorter range interaction than the two-step 3N mechanisms and are thought to be important in the 3N interaction.

In summary this first  determination of the $^{12}$C($\vec{\gamma},pd$) asymmetry shows that reactions leading to low lying states in $^{9}$Be proceed through the photon interacting with the detected nucleons in a similar manner to the  $^{3}$He($\vec{\gamma},pd$) reaction for $E_{\gamma}$ below the $\Delta$ resonance. The asymmetries at higher missing energy do not resemble  $^{3}$He($\vec{\gamma},pd$) and have a plausible explanation  in terms of multistep processes involving more than three nucleons. These new results will provide valuable  constraints on the reaction mechanisms for $(\gamma,pd)$ in heavier nuclei and their potential to be used in learning about the correlated behaviour of three nucleons in a nucleus.

\begin{acknowledgments}
This work was supported by the UK EPSRC, the British Council, the DFG (Mu 705/3, SSP 1043), BMFT (06 Tu¨ 656), DAAD (313-ARC-IX-95/41), the EC [SCI.0910.C(JR)] and NATO (CRG 970268). 
%D.G would like to thank EPSRC for research studentships during the period of this work.
\end{acknowledgments}
 %The authors would like to thank the Institut f\"{u}r Kernphysik der Universit\"{a}t Mainz for the use of its facilities and for the generous assistance provided during the course of this experiment.
\bibliography{gpdpap}

\begin{thebibliography}{36}
\expandafter\ifx\csname natexlab\endcsname\relax\def\natexlab#1{#1}\fi
\expandafter\ifx\csname bibnamefont\endcsname\relax
  \def\bibnamefont#1{#1}\fi
\expandafter\ifx\csname bibfnamefont\endcsname\relax
  \def\bibfnamefont#1{#1}\fi
\expandafter\ifx\csname citenamefont\endcsname\relax
  \def\citenamefont#1{#1}\fi
\expandafter\ifx\csname url\endcsname\relax
  \def\url#1{\texttt{#1}}\fi
\expandafter\ifx\csname urlprefix\endcsname\relax\def\urlprefix{URL }\fi
\providecommand{\bibinfo}[2]{#2}
\providecommand{\eprint}[2][]{\url{#2}}

\bibitem[{\citenamefont{Skibinski
  et~al.}(2003{\natexlab{a}})}]{Skibinski:2002ak}
\bibinfo{author}{\bibfnamefont{R.}~\bibnamefont{Skibinski}}
  \bibnamefont{et~al.}, \bibinfo{journal}{Phys. Rev.}
  \textbf{\bibinfo{volume}{C67}}, \bibinfo{pages}{054001}
  (\bibinfo{year}{2003}{\natexlab{a}}).

\bibitem[{\citenamefont{Skibinski
  et~al.}(2003{\natexlab{b}})}]{Skibinski:2003de}
\bibinfo{author}{\bibfnamefont{R.}~\bibnamefont{Skibinski}}
  \bibnamefont{et~al.}, \bibinfo{journal}{Phys. Rev.}
  \textbf{\bibinfo{volume}{C67}}, \bibinfo{pages}{054002}
  (\bibinfo{year}{2003}{\natexlab{b}}).

\bibitem[{\citenamefont{Deltuva et~al.}(2004)}]{Deltuva:2003xe}
\bibinfo{author}{\bibfnamefont{A.}~\bibnamefont{Deltuva}} \bibnamefont{et~al.},
  \bibinfo{journal}{Phys. Rev.} \textbf{\bibinfo{volume}{C69}},
  \bibinfo{pages}{034004} (\bibinfo{year}{2004}).

\bibitem[{\citenamefont{Laget}(1988)}]{Laget:1988hv}
\bibinfo{author}{\bibfnamefont{J.~M.} \bibnamefont{Laget}},
  \bibinfo{journal}{Phys. Rev.} \textbf{\bibinfo{volume}{C38}},
  \bibinfo{pages}{2993} (\bibinfo{year}{1988}).

\bibitem[{\citenamefont{Isbert et~al.}(1994)}]{Isbert:1993xf}
\bibinfo{author}{\bibfnamefont{V.}~\bibnamefont{Isbert}} \bibnamefont{et~al.},
  \bibinfo{journal}{Nucl. Phys.} \textbf{\bibinfo{volume}{A578}},
  \bibinfo{pages}{525} (\bibinfo{year}{1994}).

\bibitem[{\citenamefont{Sober et~al.}(1983)}]{gpd2}
\bibinfo{author}{\bibfnamefont{D.}~\bibnamefont{Sober}} \bibnamefont{et~al.},
  \bibinfo{journal}{Phys. Rev.} \textbf{\bibinfo{volume}{C28}},
  \bibinfo{pages}{2234} (\bibinfo{year}{1983}).

\bibitem[{\citenamefont{Argan et~al.}(1975)}]{argan}
\bibinfo{author}{\bibfnamefont{P.}~\bibnamefont{Argan}} \bibnamefont{et~al.},
  \bibinfo{journal}{Nucl. Phys.} \textbf{\bibinfo{volume}{A237}},
  \bibinfo{pages}{447} (\bibinfo{year}{1975}).

\bibitem[{\citenamefont{Gassen et~al.}(1981)}]{Gassen:1981ek}
\bibinfo{author}{\bibfnamefont{H.~J.} \bibnamefont{Gassen}}
  \bibnamefont{et~al.}, \bibinfo{journal}{Z. Phys.}
  \textbf{\bibinfo{volume}{A303}}, \bibinfo{pages}{35} (\bibinfo{year}{1981}).

\bibitem[{\citenamefont{Picozza et~al.}(1970)}]{gpd5}
\bibinfo{author}{\bibfnamefont{P.}~\bibnamefont{Picozza}} \bibnamefont{et~al.},
  \bibinfo{journal}{Nucl. Phys.} \textbf{\bibinfo{volume}{A157}},
  \bibinfo{pages}{190} (\bibinfo{year}{1970}).

\bibitem[{\citenamefont{Kolb et~al.}(1994)}]{kolb}
\bibinfo{author}{\bibfnamefont{N.}~\bibnamefont{Kolb}} \bibnamefont{et~al.},
  \bibinfo{journal}{Phys. Rev.} \textbf{\bibinfo{volume}{C49}},
  \bibinfo{pages}{2586} (\bibinfo{year}{1994}).

\bibitem[{\citenamefont{Hartmann et~al.}(1973)}]{Hartmann:1973}
\bibinfo{author}{\bibfnamefont{H.}~\bibnamefont{Hartmann}}
  \bibnamefont{et~al.}, \bibinfo{journal}{Photonuclear reactions and
  applications, Asilomar, USA, Lawrence Livermore Lab}
  \textbf{\bibinfo{volume}{Conf. no. 730301}} (\bibinfo{year}{1973}).

\bibitem[{\citenamefont{McAllister et~al.}(1999)}]{McAllister:1999mi}
\bibinfo{author}{\bibfnamefont{S.~J.} \bibnamefont{McAllister}}
  \bibnamefont{et~al.}, \bibinfo{journal}{Phys. Rev.}
  \textbf{\bibinfo{volume}{C60}}, \bibinfo{pages}{044610}
  (\bibinfo{year}{1999}).

\bibitem[{\citenamefont{MacGregor et~al.}(1998)}]{MacGregor:1998}
\bibinfo{author}{\bibfnamefont{I.~J.~D.} \bibnamefont{MacGregor}}
  \bibnamefont{et~al.}, \bibinfo{journal}{Phys. Rev. Lett.}
  \textbf{\bibinfo{volume}{80}}, \bibinfo{pages}{245} (\bibinfo{year}{1998}).

\bibitem[{\citenamefont{Branford et~al.}(2000)}]{Branford:1999cp}
\bibinfo{author}{\bibfnamefont{D.}~\bibnamefont{Branford}}
  \bibnamefont{et~al.}, \bibinfo{journal}{Phys. Rev.}
  \textbf{\bibinfo{volume}{C61}}, \bibinfo{pages}{014603}
  (\bibinfo{year}{2000}).

\bibitem[{\citenamefont{Audit et~al.}(1997)}]{Audit:1996tq}
\bibinfo{author}{\bibfnamefont{G.}~\bibnamefont{Audit}} \bibnamefont{et~al.},
  \bibinfo{journal}{Nucl. Phys.} \textbf{\bibinfo{volume}{A614}},
  \bibinfo{pages}{461} (\bibinfo{year}{1997}).

\bibitem[{\citenamefont{Watts et~al.}(2003)}]{Watts:2002jt}
\bibinfo{author}{\bibfnamefont{D.~P.} \bibnamefont{Watts}}
  \bibnamefont{et~al.}, \bibinfo{journal}{Phys. Lett.}
  \textbf{\bibinfo{volume}{B553}}, \bibinfo{pages}{25} (\bibinfo{year}{2003}).

\bibitem[{\citenamefont{Belyaev et~al.}(1984)}]{Belyaev:1985kg}
\bibinfo{author}{\bibfnamefont{A.~A.} \bibnamefont{Belyaev}}
  \bibnamefont{et~al.}, \bibinfo{journal}{JETP Lett.}
  \textbf{\bibinfo{volume}{40}}, \bibinfo{pages}{1275} (\bibinfo{year}{1984}).

\bibitem[{\citenamefont{Fabbri et~al.}(1972)}]{fabbri:1972}
\bibinfo{author}{\bibfnamefont{F.~L.} \bibnamefont{Fabbri}}
  \bibnamefont{et~al.}, \bibinfo{journal}{Lett. Nuevo Cim.}
  \textbf{\bibinfo{volume}{3}}, \bibinfo{pages}{3} (\bibinfo{year}{1972}).

\bibitem[{\citenamefont{Carrasco et~al.}(1994)\citenamefont{Carrasco,
  Vicente~Vacas, and Oset}}]{Carrasco:1992mg}
\bibinfo{author}{\bibfnamefont{R.~C.} \bibnamefont{Carrasco}},
  \bibinfo{author}{\bibfnamefont{M.~J.} \bibnamefont{Vicente~Vacas}},
  \bibnamefont{and} \bibinfo{author}{\bibfnamefont{E.}~\bibnamefont{Oset}},
  \bibinfo{journal}{Nucl. Phys.} \textbf{\bibinfo{volume}{A570}},
  \bibinfo{pages}{701} (\bibinfo{year}{1994}).

\bibitem[{\citenamefont{Ruth. et~al.}(1994)}]{ruth:1994}
\bibinfo{author}{\bibfnamefont{C.}~\bibnamefont{Ruth.}} \bibnamefont{et~al.},
  \bibinfo{journal}{Phys. Rev. Lett.} \textbf{\bibinfo{volume}{72}},
  \bibinfo{pages}{617} (\bibinfo{year}{1994}).

\bibitem[{\citenamefont{Sarty et~al.}(1993)}]{Sarty:1993}
\bibinfo{author}{\bibfnamefont{A.}~\bibnamefont{Sarty}} \bibnamefont{et~al.},
  \bibinfo{journal}{Phys. Rev.} \textbf{\bibinfo{volume}{C47}},
  \bibinfo{pages}{459} (\bibinfo{year}{1993}).

\bibitem[{\citenamefont{Herminghaus.}(1990)}]{mamib}
\bibinfo{author}{\bibfnamefont{H.}~\bibnamefont{Herminghaus.}}, in
  \emph{\bibinfo{booktitle}{Proc. of the Linear Accelerator Conference,
  Albuquerque, NM}} (\bibinfo{year}{1990}).

\bibitem[{\citenamefont{Walcher}(1990)}]{mamib2}
\bibinfo{author}{\bibfnamefont{T.}~\bibnamefont{Walcher}},
  \bibinfo{journal}{Prog. Part. Nucl. Phys.} \textbf{\bibinfo{volume}{24}},
  \bibinfo{pages}{189} (\bibinfo{year}{1990}).

\bibitem[{\citenamefont{Anthony et~al.}(1985)}]{Anthony:1986ge}
\bibinfo{author}{\bibfnamefont{I.}~\bibnamefont{Anthony}} \bibnamefont{et~al.},
  \bibinfo{journal}{Nucl. Phys.} \textbf{\bibinfo{volume}{A446}},
  \bibinfo{pages}{322c} (\bibinfo{year}{1985}).

\bibitem[{\citenamefont{Hall et~al.}(1996)}]{Hall:1996gh}
\bibinfo{author}{\bibfnamefont{S.~J.} \bibnamefont{Hall}} \bibnamefont{et~al.},
  \bibinfo{journal}{Nucl. Instrum. Meth.} \textbf{\bibinfo{volume}{A368}},
  \bibinfo{pages}{698} (\bibinfo{year}{1996}).

\bibitem[{\citenamefont{MacGregor et~al.}(1996)}]{MacGregor:1996tb}
\bibinfo{author}{\bibfnamefont{I.~J.~D.} \bibnamefont{MacGregor}}
  \bibnamefont{et~al.}, \bibinfo{journal}{Nucl. Instrum. Meth.}
  \textbf{\bibinfo{volume}{A382}}, \bibinfo{pages}{479} (\bibinfo{year}{1996}).

\bibitem[{\citenamefont{Grabmayr et~al.}(1998)}]{Grabmayr:1998ez}
\bibinfo{author}{\bibfnamefont{P.}~\bibnamefont{Grabmayr}}
  \bibnamefont{et~al.}, \bibinfo{journal}{Nucl. Instrum. Meth.}
  \textbf{\bibinfo{volume}{A402}}, \bibinfo{pages}{85} (\bibinfo{year}{1998}).

\bibitem[{\citenamefont{Powrie et~al.}(2001)}]{Powrie:2001af}
\bibinfo{author}{\bibfnamefont{C.~J.~Y.} \bibnamefont{Powrie}}
  \bibnamefont{et~al.}, \bibinfo{journal}{Phys. Rev.}
  \textbf{\bibinfo{volume}{C64}}, \bibinfo{pages}{034602}
  (\bibinfo{year}{2001}).

\bibitem[{\citenamefont{Franczuk et~al.}(1999)}]{Franczuk:1999}
\bibinfo{author}{\bibfnamefont{S.}~\bibnamefont{Franczuk}}
  \bibnamefont{et~al.}, \bibinfo{journal}{Phys. Lett.}
  \textbf{\bibinfo{volume}{B450}}, \bibinfo{pages}{332} (\bibinfo{year}{1999}).

\bibitem[{\citenamefont{Lohmann et~al.}(1994)}]{Lohmann:1994vz}
\bibinfo{author}{\bibfnamefont{D.}~\bibnamefont{Lohmann}} \bibnamefont{et~al.},
  \bibinfo{journal}{Nucl. Instrum. Meth.} \textbf{\bibinfo{volume}{A343}},
  \bibinfo{pages}{494} (\bibinfo{year}{1994}).

\bibitem[{\citenamefont{Kraus et~al.}(1997)}]{Kraus:1997pp}
\bibinfo{author}{\bibfnamefont{A.}~\bibnamefont{Kraus}} \bibnamefont{et~al.},
  \bibinfo{journal}{Phys. Rev. Lett.} \textbf{\bibinfo{volume}{79}},
  \bibinfo{pages}{3834} (\bibinfo{year}{1997}).

\bibitem[{\citenamefont{Rambo et~al.}(1998)}]{Rambo:1998iu}
\bibinfo{author}{\bibfnamefont{F.}~\bibnamefont{Rambo}} \bibnamefont{et~al.},
  \bibinfo{journal}{Phys. Rev.} \textbf{\bibinfo{volume}{C58}},
  \bibinfo{pages}{489} (\bibinfo{year}{1998}).

\bibitem[{\citenamefont{Natter et~al.}(2002)}]{Natter:2002}
\bibinfo{author}{\bibfnamefont{F.}~\bibnamefont{Natter}} \bibnamefont{et~al.},
  \bibinfo{journal}{Nucl. Instrum. Meth.} \textbf{\bibinfo{volume}{A481}}
  (\bibinfo{year}{2002}).

\bibitem[{\citenamefont{Lamparter et~al.}(1996)}]{Lamparter:1996va}
\bibinfo{author}{\bibfnamefont{T.}~\bibnamefont{Lamparter}}
  \bibnamefont{et~al.}, \bibinfo{journal}{Z. Phys.}
  \textbf{\bibinfo{volume}{A355}}, \bibinfo{pages}{1} (\bibinfo{year}{1996}).

\bibitem[{\citenamefont{Watts et~al.}(2000)}]{Watts:2000dk}
\bibinfo{author}{\bibfnamefont{D.~P.} \bibnamefont{Watts}}
  \bibnamefont{et~al.}, \bibinfo{journal}{Phys. Rev.}
  \textbf{\bibinfo{volume}{C62}}, \bibinfo{pages}{014616}
  (\bibinfo{year}{2000}).

\bibitem[{\citenamefont{Dreschsel et~al.}(1999)\citenamefont{Dreschsel,
  Kamalov, and Tiator}}]{maid}
\bibinfo{author}{\bibfnamefont{D.}~\bibnamefont{Dreschsel}},
  \bibinfo{author}{\bibfnamefont{S.}~\bibnamefont{Kamalov}}, \bibnamefont{and}
  \bibinfo{author}{\bibfnamefont{L.}~\bibnamefont{Tiator}},
  \bibinfo{journal}{Nucl. Phys.} \textbf{\bibinfo{volume}{A645}},
  \bibinfo{pages}{145} (\bibinfo{year}{1999}).

\end{thebibliography}

\end{document}